\documentclass[useAMS,usenatbib]{mn2e}
\usepackage{graphics}
\usepackage{epsf}
\usepackage{subfigure}

\begin{document}

\title[Galactic PL relations]{Period-luminosity relations for Galactic Cepheid variables with independent distance measurements.}
\author[Ngeow \& Kanbur]{Chow-Choong Ngeow$^{1}$\thanks{E-mail: ngeow@nova.astro.umass.edu} and Shashi M. Kanbur$^{1}$ 
\\
$^{1}$Department of Astronomy, University of Massachusetts      
\\
Amherst, MA 01003, USA}


\date{Accepted 2003 month day. Received 2003 month day; in original form 2003 month day}


\maketitle

\begin{abstract}
        In this paper, we derive the period-luminosity (PL) relation for Galactic Cepheids with recent independent distance measurements from open cluster, Barnes-Evans surface brightness, interferometry and {\it HST} astrometry techniques. Our PL relation confirms the results of Tammann et al. (2003), which showed that the Galactic Cepheids follow a different PL relation to their LMC counterparts. Our results also show that the slope of the Galactic PL relation is inconsistent with the LMC slope with more than 95\% confidence level. We apply this Galactic PL relation to find the distance to NGC 4258. Our result of $\mu_o=29.49\pm0.06$(random error)$mag.$ agrees at the $\sim1.4\sigma$ level with the geometrical distance of $\mu_{geo}=29.28\pm0.15mag.$ from water maser measurements.
\end{abstract}



\section{Introduction}

        Recently, \citet[][hereafter T03]{tam03} derived the Galactic PL relation by combining the Galactic Cepheids with independent distance measurements from open clusters/associations \citep{fea99} and from the Barnes-Evans (BE) surface brightness techniques \citep{gie98}. The resulting Galactic PL relations in T03 are steeper than the LMC PL relations commonly applied in distance scale applications (as in, e.g., \citealt{fre01}). Similar conclusions are also reported in \citet{fou03}.

        The need to use the Galactic PL relation as a fundamental calibrating relation has become more desirable in recent years \citep{fea03,fou03,kan03,tam03,thi03}, because of the following two main reasons: (a) The average value of metallicity (defined as $12+\log[O/H]$) in target galaxies of the $H_0$ Key Project is $\sim8.84\pm0.31dex$ \citep{fre01}, which is closer to the standard Solar value of $8.87\pm0.07dex$ \citep{gre96} than the LMC value of $8.50\pm0.08dex$ (see reference in \citealt{fer00}); and (b) There is evidence that the LMC PL relation is broken at 10 days \citep{tam02,kan03b}, i.e., the short ($<10$ days) and long period Cepheids in the LMC follow different PL relations. Due to these reasons, the calibrated Galactic PL relation will become important in future distance scale studies.

        In this paper, we derive the Galactic PL relation from Cepheids with independent distance measurements. Our analysis of the Galactic PL relation is similar to T03 but different in the following aspects:

        \begin{enumerate}
        \item   In addition to the Cepheids from \citet{fea99} and \citet{gie98} that are used in T03, we include other recent distance measurements to Galactic Cepheids that are available in the literature (see Section 2). These include 11 additional Cepheids that are not included in T03.
        \item   As most of the Cepheids we considered here have more than one independent distance measurement, we took the standard weighted-mean of the available distances as the final adopted distance to derive the PL relation. 
        \end{enumerate}


\section{Galactic Cepheids with independent distance measurement}

        We collected the Galactic Cepheids with recent distance measurements from the literature, which include:
        
        \begin{enumerate}
        \item   {\bf Distances from Open Cluster techniques}: These distances are adopted from table 3 of T03, where the authors adopted a distance modulus for the Pleiades of $\mu_{Pleiades}=5.61\pm0.03mag.$ \citep{ste01}. \citet{fea99} estimated that the uncertainty associated with cluster distance moduli is $\sim0.15$-$0.20mag.$ (see, e.g., \citealt{rom89}), hence we assign an uncertainty of $0.20mag.$ to the open cluster distance moduli ($\mu_o[O.C.]$) in Table \ref{tab1}. This error could incorporate the uncertainty due to the location of the Cepheid in the cluster (far-side vs. near-side), the uncertainty due to metallicity corrections from cluster to cluster, etc. (see also \citealt{tur02}). In addition, we also include other open cluster distances from \citet{tur02} and \citet{hoy03} in Table \ref{tab1}. Since these additional open cluster distances are based on $\mu_{Pleiades}=5.56mag.$, we add a correction of $+0.05mag$ to these distances\footnote{Thanks to G. A. Tammann for pointing out this correction to \citet{hoy03} distances.} in Table \ref{tab1} to be consistent with the T03 open cluster distances. However, it is unclear whether the distances given in \citet[][where the T03 adopted distances originate]{fea99} and in \citet{tur02} are totally independent to each other or not, since some of them were adopted from the same sources. For example, the open cluster distances for CV MON in \citet{fea99} and \citet{tur02} are all adopted from \citet{tur98}. We have labelled the \citet{tur02} distances that are dubious in this dependency regard in Table \ref{tab1}, and excluded them in obtaining the weighted-mean distances. For GT CAR, CG CAS and TV CMa in \citet{tur02}, which are not included in \citet{fea99}, there is no I band data available in the literature. We exclude them in order to have a consistent number of Cepheids in B, V and I bands.

        \item   {\bf Distances from BE surface brightness techniques}: We adopt these distances from \citet{fou03}, which is an updated version of \citet{gie98}, with additional Cepheids that are not included in \citet{gie98}. Fouqu\'{e} (2003, private communication) has verified that the the distances from \citet{gie98} and \citet{fou03} are not independent of each other, as the distances from the later paper use the latest available data to update the distances in \citet{gie98}. The exception is CS VEL, where we include the distance from \citet{gie98} since there is no distance given to this Cepheid in \citet{fou03}. Note that T03 only included the distances from \citet{gie98} but not from \citet{fou03}. We also include the latest distance measurements using BE techniques from \citet{bar03}, who used a Bayesian statistical approach in their analysis. 

        \item   {\bf Distances from interferometry techniques}: These distances are adopted from \citet{lan02}, \citet{nor02} and \cite{ker03}. Note that for distances in \citet{ker03}, we did not use the direct distance measurements for the four Cepheids ($\eta$ AQL, W SGR, $\beta$ DOR \& $l$ CAR) because the errors are asymmetric and large (except for $l$ CAR). Instead we adopt the distances from their table 12. These are obtained from combining the interferometry data and the empirical Period-Radius relations from \citet{gie98}.

        \item   {\bf Distance from {\it HST} astrometry techniques}: Currently there is only one Cepheid with a distance measurement from astrometry using the {\it Hubble Space Telescope}: $\delta$ CEP from \citet{ben02}. 
        \end{enumerate}

        The selected Galactic Cepheids and the corresponding distances from these sources are summarized in Table \ref{tab1}. Since these distances are from independent measurements, we can take the weighted-mean among the available distances (i.e. column 2, 3, 5 \& 6 in Table \ref{tab1}). Note that T03 only uses the distance moduli in column 2 \& 4 of Table \ref{tab1} in their paper. We did not include the distances from {\it Hipparcos} because the errors in the distance moduli, after conversion from parallax, are large (see, e.g., \citealt{mad98}).

        In this work, we exclude all the possible non-fundamental mode Cepheids and those not classified as ``DCEP'' in the General Catalogue of Variable Stars (GCVS, \citealt{kho98}). Some of them are mentioned in T03, and include: EV SCT, V1726 CYG, SZ TAU, QZ NOR, $\alpha$ UMi and V367 SCT. We further exclude EU TAU from \citet{bar03} because it is a first-overtone Cepheid. Note that LS PUP is not classified as ``DCEP'' in the GCVS. We also exclude this Cepheid though it is included in the T03 sample. For Cepheids in \citet{tur02} and \citet{ker03}, we did not include the Cepheids that are classified as ``DCEPS'' in the GCVS, which includes SU CAS, GH CAR and Y OPH. The additional Cepheids included in our sample but not in T03 are labelled in Table \ref{tab1}. 

        In principle the number of Cepheids can be increased if we include the first overtone (FO) Galactic Cepheids in our sample, by using their fundamental mode periods in obtaining the PL relation. However, we prefer not to include the FO Cepheids due to the following reasons: (a) we want to avoid the contaminations from other types of Cepheids and select only the {\it bona fide} fundamental mode Cepheids; (b) we want to be consistent with T03, who excluded the non-fundamental mode Cepheids in their study; (c) the physics involved in fundamental mode and FO Cepheids is not the same (see, e.g., \citealt{ant94,ant95,bon99,feu00,bon02}) and (d) the results of microlensing surveys to the Magellanic Clouds suggest that the FO Cepheids follow their own PL relations (see, e.g., \citealt{uda99}). The theoretical studies from \citet{bon99b} and \citet{bar01} also suggest that the FO Cepheids follow different PL relations. Nevertheless, we list the FO Cepheids (classified as ``DCEPS'' in the GCVS) with recent distance measurements in Table \ref{FO} for completeness. Note that although BD CAS has been updated to ``DCEPS'' by \citet{por94}, there is no I band data available for this Cepheid in the literature. We therefore exclude BD CAS in Table \ref{FO}.


        \begin{table*}
        \centering
        \caption{Distances to Galactic fundamental mode Cepheids.}
        \label{tab1}
        \begin{tabular}{lcccclc} \hline
        Cepheid & $\mu_o(O.C.)^{\mathrm a}$     & $\mu_o(TB)^{\mathrm a}$ & $\mu_o(G98)^{\mathrm a}$    & $\mu_o(F03)^{\mathrm a}$         &$\mu_o(other)^{\mathrm a}$           & $\mu_o(w.m.)^{\mathrm a}$ \\ 
        (1)     & (2) & (3) & (4) & (5) & (6)  & (7) \\
        \hline \hline
        $\eta$ AQL$^{\mathrm i}$        & $\cdots$      & $\cdots$                      & $\cdots$                      & $6.986\pm0.052$  & $7.526\pm0.217^{\mathrm b}$; $7.108\pm0.148^{\mathrm e}$   & $7.025\pm0.048$ \\
        RX AUR$^{\mathrm i}$            & $\cdots$      & $\cdots$                      & $\cdots$                      & $\cdots$         & $11.101\pm0.204^{\mathrm c}$                               & $11.101\pm0.204$\\
        U   CAR                         & $11.46$       & $11.46\pm0.04^{\mathrm g}$    & $11.069\pm0.038$              & $10.972\pm0.032$ & $\cdots$                                                   & $10.984\pm0.032$\\
        VY  CAR                         & $11.63$       & $11.60\pm0.09^{\mathrm g}$    & $11.419\pm0.043$              & $11.501\pm0.022$ & $\cdots$                                                   & $11.503\pm0.022$\\
        WZ  CAR                         & $\cdots$      & $\cdots$                      & $12.980\pm0.135$              & $12.918\pm0.066$ & $\cdots$                                                   & $12.918\pm0.066$\\
        $l$ CAR                         & $\cdots$      & $\cdots$                      & $8.941\pm0.053$               & $8.989\pm0.032$  & $8.670\pm0.204^{\mathrm e}$                                & $8.981\pm0.032$ \\
        CEa CAS                         & $12.69$       & $12.74\pm0.15$                & $\cdots$                      & $\cdots$         & $12.63\pm0.14^{\mathrm d}$                                 & $12.662\pm0.091$\\
        CEb CAS                         & $12.69$       & $12.74\pm0.15$                & $\cdots$                      & $\cdots$         & $12.63\pm0.14^{\mathrm d}$                                 & $12.662\pm0.091$\\
        CF  CAS                         & $12.69$       & $12.74\pm0.15$                & $\cdots$                      & $\cdots$         & $12.63\pm0.14^{\mathrm d}$                                 & $12.662\pm0.091$\\
        DL  CAS                         & $11.22$       & $11.16\pm0.04^{\mathrm g}$    & $\cdots$                      & $\cdots$         & $10.99\pm0.14^{\mathrm d}$                                 & $11.032\pm0.115$\\
        V   CEN                         & $9.17$        & $9.16\pm0.04^{\mathrm g}$     & $9.302\pm0.024$               & $9.175\pm0.063$  & $\cdots$                                                   & $9.175\pm0.060$ \\
        VW  CEN                         & $\cdots$      & $\cdots$                      & $13.014\pm0.042$              & $12.803\pm0.039$ & $\cdots$                                                   & $12.803\pm0.039$\\
        XX  CEN                         & $\cdots$      & $\cdots$                      & $10.847\pm0.065$              & $11.116\pm0.023$ & $\cdots$                                                   & $11.116\pm0.023$\\
        KN  CEN                         & $\cdots$      & $\cdots$                      & $12.911\pm0.060$              & $13.124\pm0.045$ & $\cdots$                                                   & $13.124\pm0.045$\\
        $\delta$ CEP$^{\mathrm i}$      & $\cdots$      & $6.76\pm0.10$                 & $\cdots$                      & $7.084\pm0.044$  & $7.173\pm0.048^{\mathrm b}$; $7.181\pm0.089^{\mathrm f}$   & $7.100\pm0.029$ \\
        X   CYG$^{\mathrm i}$           & $\cdots$      & $10.30\pm0.05$                & $\cdots$                      & $10.421\pm0.016$ & $10.209\pm0.055^{\mathrm c}$                               & $10.395\pm0.015$\\
        SU CYG$^{\mathrm i}$            & $\cdots$      & $9.70\pm0.06$                 & $\cdots$                      & $\cdots$         & $\cdots$                                                   & $9.700\pm0.060$ \\      
        $\beta$ DOR$^{\mathrm i}$       & $\cdots$      & $\cdots$                      & $\cdots$                      & $\cdots$         & $7.566\pm0.153^{\mathrm e}$                                & $7.566\pm0.153$ \\
        $\zeta$ GEM$^{\mathrm i}$       & $\cdots$      & $7.85\pm0.10$                 & $\cdots$                      & $\cdots$         & $7.794\pm0.228^{\mathrm b}$; $7.782\pm0.211^{\mathrm e}$   & $7.732\pm0.084$ \\
        Z   LAC$^{\mathrm i}$           & $\cdots$      & $\cdots$                      & $\cdots$                      & $11.637\pm0.055$ & $\cdots$                                                   & $11.637\pm0.055$\\ 
        T   MON                         & $11.14$       & $11.05\pm0.14^{\mathrm g}$    & $10.576\pm0.067$              & $10.777\pm0.053$ & $10.580\pm0.068^{\mathrm c}$                               & $10.721\pm0.041$\\
        CV  MON                         & $11.22$       & $11.12\pm0.04^{\mathrm g}$    & $10.901\pm0.046$              & $10.988\pm0.034$ & $11.39\pm0.21^{\mathrm d}$                                 & $11.003\pm0.033$\\
        UU  MUS                         & $\cdots$      & $\cdots$                      & $12.260\pm0.092$              & $12.589\pm0.084$ & $\cdots$                                                   & $12.589\pm0.084$\\
        S   NOR                         & $9.85$        & $9.82\pm0.04^{\mathrm g}$     & $9.918\pm0.025$               & $9.908\pm0.032$  & $\cdots$                                                   & $9.907\pm0.032$ \\
        U   NOR                         & $\cdots$      & $\cdots$                      & $10.769\pm0.067$              & $10.716\pm0.060$ & $\cdots$                                                   & $10.716\pm0.060$\\
        TW  NOR                         & $11.47$       & $11.47\pm0.08^{\mathrm g}$    & $\cdots$                      & $\cdots$         & $11.38\pm0.18^{\mathrm d}$                                 & $11.393\pm0.134$\\
        V340 NOR                        & $11.17$       & $11.16\pm0.11$                & $11.498\pm0.130$              & $11.145\pm0.185$ & $11.23\pm0.12^{\mathrm d}$                                 & $11.166\pm0.070$\\
        BF  OPH                         & $\cdots$      & $\cdots$                      & $9.496\pm0.110$               & $9.271\pm0.034$  & $9.265\pm0.192^{\mathrm c}$                                & $9.271\pm0.033$ \\
        UY  PER                         & $11.78$       & $11.88\pm0.50^{\mathrm g}$    & $\cdots$                      & $\cdots$         & $\cdots$                                                   & $11.780\pm0.200$\\
        RS  PUP                         & $11.28$       & $\cdots$                      & $\cdots$                      & $11.622\pm0.076$ & $11.160\pm0.290^{\mathrm c}$                               & $11.555\pm0.069$\\
        VZ  PUP                         & $\cdots$      & $\cdots$                      & $13.551\pm0.036$              & $13.083\pm0.057$ & $\cdots$                                                   & $13.083\pm0.057$\\
        AQ  PUP                         & $\cdots$      & $11.78\pm0.10$                & $12.750\pm0.038$              & $12.522\pm0.045$ & $\cdots$                                                   & $12.397\pm0.041$\\
        BN  PUP                         & $\cdots$      & $\cdots$                      & $12.924\pm0.051$              & $12.950\pm0.050$ & $\cdots$                                                   & $12.950\pm0.050$\\
        GY  SGE                         & $12.65$       & $12.74\pm0.08^{\mathrm g}$    & $12.939\pm0.071^{\mathrm h}$  & $\cdots$         & $\cdots$                                                   & $12.650\pm0.200$\\
        U   SGR                         & $9.07$        & $8.94\pm0.10^{\mathrm g}$     & $8.869\pm0.015$               & $8.871\pm0.022$  & $9.137\pm0.158^{\mathrm c};9.13\pm0.18^{\mathrm d}$        & $8.881\pm0.022$ \\
        W   SGR$^{\mathrm i}$           & $\cdots$      & $\cdots$                      & $\cdots$                      & $\cdots$         & $7.933\pm0.169^{\mathrm e}$                                & $7.933\pm0.169$ \\
        X   SGR$^{\mathrm i}$           & $\cdots$      & $\cdots$                      & $\cdots$                      & $\cdots$         & $7.553\pm0.161^{\mathrm e}$                                & $7.553\pm0.161$ \\
        WZ  SGR                         & $11.26$       & $11.31\pm0.04$                & $11.262\pm0.021$              & $11.287\pm0.047$ & $12.001\pm0.169^{\mathrm c};11.23\pm0.16^{\mathrm d}$      & $11.316\pm0.029$\\
        BB  SGR                         & $9.11$        & $9.08\pm0.08^{\mathrm g}$     & $9.238\pm0.022$               & $9.519\pm0.028$  & $9.805\pm0.181^{\mathrm c}$                                & $9.518\pm0.027$ \\
        RY  SCO                         & $\cdots$      & $\cdots$                      & $10.469\pm0.042$              & $10.516\pm0.034$ & $9.911\pm0.147^{\mathrm c}$                                & $10.485\pm0.033$\\
        KQ  SCO                         & $12.36$       & $12.33\pm0.25^{\mathrm g}$    & $\cdots$                      & $\cdots$         & $\cdots$                                                   & $12.360\pm0.200$\\
        RU  SCT                         & $11.60$       & $11.64\pm0.14^{\mathrm g}$    & $\cdots$                      & $\cdots$         & $11.41\pm0.20^{\mathrm d}$                                 & $11.480\pm0.141$\\
        T   VEL                         & $\cdots$      & $\cdots$                      & $10.094\pm0.023$              & $9.802\pm0.060$  & $\cdots$                                                   & $9.802\pm0.060$ \\    
        RY  VEL                         & $\cdots$      & $\cdots$                      & $12.100\pm0.050$              & $12.019\pm0.032$ & $\cdots$                                                   & $12.019\pm0.032$\\
        RZ  VEL                         & $11.19$       & $11.27\pm0.31^{\mathrm g}$    & $11.169\pm0.025$              & $11.020\pm0.029$ & $\cdots$                                                   & $11.024\pm0.029$\\
        SW  VEL                         & $12.08$       & $12.04\pm0.05^{\mathrm g}$    & $11.989\pm0.056$              & $11.998\pm0.025$ & $\cdots$                                                   & $11.999\pm0.025$\\
        CS  VEL                         & $12.59$       & $12.59\pm0.14^{\mathrm g}$    & $12.713\pm0.144$              & $\cdots$         & $\cdots$                                                   & $12.671\pm0.117$\\
        S   VUL                         & $13.24$       & $13.24\pm0.09^{\mathrm g}$    & $13.731\pm0.095^{\mathrm h}$  & $\cdots$         & $\cdots$                                                   & $13.240\pm0.200$\\    
        T   VUL$^{\mathrm i}$           & $\cdots$      & $\cdots$                      & $\cdots$                      & $\cdots$         & $8.920\pm0.146^{\mathrm c}$                                & $8.920\pm0.146$ \\
        SV  VUL                         & $11.83$       & $11.78\pm0.05$                & $12.325\pm0.072^{\mathrm h}$  & $\cdots$         & $11.331\pm0.081^{\mathrm c};10.98\pm0.21^{\mathrm d}$      & $11.636\pm0.041$\\    
        \hline
        \end{tabular}
        \begin{list}{}{}
        \item   $^{\mathrm a}$ $\mu_o(O.C.)$ = open cluster distance from T03; $\mu_o{(TB)}$ = open cluster distance from \citet{tur02}, adjusted to $\mu_{Pleiades}=5.61mag.$ by adding $+0.05mag$.; $\mu_o{(G98)}$ = BE distance from \citet{gie98}; $\mu_o{(F03)}$ = BE distance from \citet{fou03}; $\mu_o{(other)}$ = distance from other sources; $\mu_o{(w.m.)}$ = the weighted-mean distance for the entries in column 2, 3, 5 \& 6, when available. Excepts for CS VEL, which we include the $\mu_o(G98)$ in obtaining the weighted-mean. 
        \item   $^{\mathrm b}$ Distance measurements from interferometry: $\eta$ AQL \& $\zeta$ GEM are from \citet{lan02}, and $\delta$ CEP is from \citet{nor02}. 
        \item   $^{\mathrm c}$ BE distance measurements from \citet{bar03}, who use a Bayesian approach in their analysis. 
        \item   $^{\mathrm d}$ Open cluster distances from \citet{hoy03}, adjusted to $\mu_{Pleiades}=5.61mag.$ by adding $+0.05mag$. 
        \item   $^{\mathrm e}$ Interferometry distance measurements from \citet{ker03}. 
        \item   $^{\mathrm f}$ {\it HST} astrometric measurement from \citet{ben02}. 
        \item   $^{\mathrm g}$ These distances are not included in calculating the weighted-mean distances, see text for details.
        \item   $^{\mathrm h}$ These distance moduli are not used in both \citet{gie98} and T03 as they appear to be outliers in the PL plots.  
        \item   $^{\mathrm i}$ These Cepheids are not included in T03. 
        \end{list}
        \end{table*}


        \begin{table*}
        \centering
        \caption{Distances to Galactic first overtone Cepheids.}
        \label{FO}
        \begin{tabular}{lccccccc} \hline
        Cepheid & $\mu_o(O.C.)^{\mathrm a}$ & $\mu_o(TB)^{\mathrm a}$ & $\mu_o(G98)^{\mathrm a}$ & $\mu_o(B03)^{\mathrm a}$ & $\mu_o(H03)^{\mathrm a}$& $\mu_o(K03)^{\mathrm a}$ & $\mu_o(w.m.)^{\mathrm a}$\\ 
        \hline \hline
        GH CAR          & $\cdots$           & $10.99\pm0.20$             & $\cdots$         & $\cdots$      & $\cdots$       & $\cdots$      & $10.990\pm0.200$ \\
        SU CAS          & $7.12^{\mathrm c}$ & $7.11\pm0.03^{\mathrm b}$  & $\cdots$         & $\cdots$      & $\cdots$       & $\cdots$      & $7.120\pm0.200$\\
        V1726 CYG       & 11.02              & $11.02\pm0.03^{\mathrm b}$ & $\cdots$         & $\cdots$      & $\cdots$       & $\cdots$      & $11.020\pm0.200$ \\
        QZ NOR          & 11.17              & $11.16\pm0.11$             & $11.095\pm0.031$ & $\cdots$      & $11.23\pm0.12$ & $\cdots$      & $11.169\pm0.075$  \\
        Y OPH           & $\cdots$           & $\cdots$                   & $\cdots$         & $\cdots$      & $\cdots$       & $9.06\pm0.21$ & $9.060\pm0.210$ \\
        EV SCT          & 10.92              & $11.09\pm0.07$             & $11.066\pm0.033$ & $\cdots$      & $10.84\pm0.15$ & $\cdots$      & $11.026\pm0.060$ \\
        SZ TAU          & 8.72               & $8.71\pm0.02^{\mathrm b}$  & $8.090\pm0.042$  & $8.74\pm0.33$ & $\cdots$       & $\cdots$      & $8.725\pm0.171$ \\
        EU TAU          & $\cdots$           & $\cdots$                   & $\cdots$         & $10.27\pm0.16$& $\cdots$       & $\cdots$      & $10.270\pm0.160$ \\
        $\alpha$ UMi    & 5.19               & $5.44\pm0.05$              & $\cdots$         & $\cdots$      & $\cdots$       & $\cdots$      & $5.425\pm0.049$ \\
        \hline
        \end{tabular}
        \begin{list}{}{}
        \item   $^{\mathrm a}$ $\mu_o(O.C.)$ = open cluster distance from T03; $\mu_o(TB)$ = open cluster from \citet{tur02}, adjusted to $\mu_{Pleiades}=5.61mag.$ by adding $+0.05mag$.; $\mu_o{(G98)}$ = BE distance from \citet{fou03}; $\mu_o(B03)$ = BE distance from \citet{bar03}; $\mu_o{(H03)}$ = open cluster distance from \citet{hoy03}, adjusted to $\mu_{Pleiades}=5.61mag.$ by adding $+0.05mag$.; $\mu_o{(K03)}$ = interferometry distance from \citet{ker03}; $\mu_o(w.m.)$ = weighted-mean distances.
        \item   $^{\mathrm b}$ These distances are not included in calculating the weighted-mean distances, since it is unclear whether these distances are independent to the distances given in \citet{fea99} or T03. 
        \item   $^{\mathrm c}$ This distance is adopted from \citet{fea99}, adjusting to $\mu_{Pleiades}=5.61mag.$ by adding $+0.04mag$.
        \end{list}
        \end{table*}


\section{The period-luminosity relation}
        
\begin{figure}
        \hbox{\hspace{0.1cm}\epsfxsize=8.0cm \epsfbox{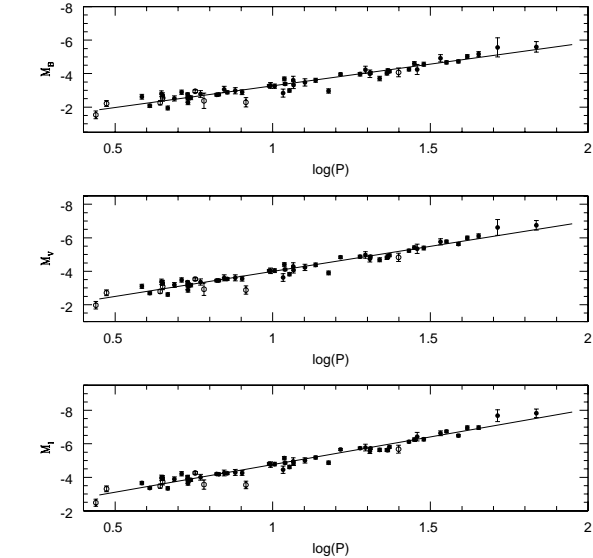}}
        \caption{Galactic PL relation in B (top panel), V (middle panel) and I (bottom panel) band. The lines are the fitted PL relations, as given in equation (1)-(3). The solid circles are for fundamental mode Cepheids as listed in Table \ref{tab1}. The open circles are the first overtone Cepheids from Table \ref{FO}, plotted with their fundamental mode periods. The error bars include the errors in distance modulus, errors in extinction and an error estimation of $0.05mag.$ in mean magnitudes.} 
        \label{pl}
        \end{figure}

        The values of $\log(P)$, $(B-V)_o$ color, $E(B-V)$\footnote{$E(B-V)_{corr}$ in T03.}, and the mean B, V and I band\footnote{There are no mean I band values for $\eta$ AQL, $\delta$ CEP and $\zeta$ GEM in T03, hence we adopted the I band mean magnitudes from \citet{lan99}.} magnitudes for the Cepheids in Table \ref{tab1} are all taken from T03. The absorption-to-reddening coefficient, $R$, for individual Cepheids is derived using the prescription given in T03\footnote{There are other formulae for the $R_V$ available in the literature. We choose the formula of $R_V$ from T03 in order to be consistent with the work of T03. The detailed study of the sensitivity of PL relation to the selected $R_V$ will be addressed in a future paper.}: $R_V = 3.17(\pm0.13)+ 0.44[(B-V)_o - 0.78] + 0.05 [E(B-V) - 0.42]$, $R_B=R_V+1.00$ and $R_I=R_V-1.28$. The B, V and I band extinction-corrected absolute magnitudes for the 50 Cepheids in Table \ref{tab1} can be calculated by adopting the weighted-mean distance moduli, as given in the last column of Table \ref{tab1}, and then fitted with least-square regressions to obtain the PL relation. The plots of the fitted Galactic PL relations are presented as solid lines in Figure \ref{pl}, with the following expressions:

        \begin{eqnarray}
        M_B=  -2.594(\pm0.106)\log(P) - 0.674(\pm0.123), \sigma=0.248 \\
        M_V=  -2.999(\pm0.097)\log(P) - 0.995(\pm0.112), \sigma=0.226 \\
        M_I\ =  -3.303(\pm0.094)\log(P) - 1.450(\pm0.108), \sigma=0.219 
        \end{eqnarray}

        The error bars in Figure \ref{pl} are obtained from the quadrature sum of the error estimates of distance modulus (given in the last column of Table \ref{tab1}), extinction (adopted from \citet{fer95} database\footnote{{\tt http://ddo.astro.utoronto.ca/cepheids.html}}) and mean magnitudes. We assign a conservative error of $0.05mag.$ to the mean magnitudes. This is reasonable because the mean magnitudes are derived from accurate and reliable light curves \citep{ber00}.

        


\subsection{The effect of FO Cepheids}

        In this section we investigate the effect of including the Galactic PL relations if we include the FO Cepheids that are listed in Table \ref{FO}, as suggested by anonymous referee. The fundamental mode periods and the mean magnitudes in B, V and I band for these Cepheids are taken from \citet{ber00}. The color excess for these Cepheids are calculated via the prescription given in T03: $E(B-V)=0.951E(B-V)_F$, where $E(B-V)_F$ is taken from \citet{fer95}. Then the color, $(B-V)_o$, can be calculated via $(B-V)-E(B-V)$. The extinction corrections for these Cepheids are handled in the same way as in fundamental mode Cepheids. The fitted PL relations from the combination of 59 fundamental mode and first overtone Cepheids are:

        \begin{eqnarray}
        \nonumber
        M_B=  -2.590(\pm0.100)\log(P) - 0.665(\pm0.111), \sigma=0.265 \\
        \nonumber
        M_V=  -3.007(\pm0.097)\log(P) - 0.962(\pm0.108), \sigma=0.257 \\
        \nonumber
        M_I\ =  -3.319(\pm0.098)\log(P) - 1.406(\pm0.108), \sigma=0.260 
        \end{eqnarray}

        By comparing these PL relations to equations (1)-(3), we see that including the FO Cepheids does not significantly alter or improve the PL relations. However, the dispersions of the PL relations ($\sigma$) have become larger than the PL relations without FO Cepheids. This is mainly due to the one outlier, GH CAR (with $\log(P_o)\sim0.91$), as shown in Figure \ref{pl}. After removing this Cepheid, the dispersions of the PL relations become comparable to those given in equations (1)-(3). Even though the PL relations with the FO Cepheids are almost identical to the PL relations without FO Cepheids, we prefer the solutions given in equations (1)-(3) to be the calibrated Galactic PL relations, as we have argued in Section 2. 

\subsection{The effect of open cluster distances from Turner \& Burke 2002}

        Since it is unclear whether some of the open cluster distances given in \citet{tur02} are totally independent of the distances given in \citet{fea99} or not, we examine the changes of PL relations if we either exclude {\it all} open cluster distances from \citet{tur02} or assume that these distances are totally independent of \citet{fea99}, and include them in obtaining the weighted-mean distances. Recall that equations (1)-(3) used some of the \citet{tur02} distances that are either excluded in or independent of \citet{fea99}. For the former case, the PL relations with 49 fundamental mode Cepheids are:
        
        \begin{eqnarray}
        \nonumber
        M_B=  -2.627(\pm0.105)\log(P) - 0.623(\pm0.123), \sigma=0.240 \\
        \nonumber
        M_V=  -3.025(\pm0.096)\log(P) - 0.954(\pm0.112), \sigma=0.220 \\
        \nonumber
        M_I\ =  -3.320(\pm0.095)\log(P) - 1.420(\pm0.110), \sigma=0.216 
        \end{eqnarray}

        For the latter case where we include {\it all} the open cluster distances from \citet{tur02}, the fitted PL relations for the 50 Cepheids in Table \ref{tab1} become:

        \begin{eqnarray}
        \nonumber
        M_B=  -2.619(\pm0.107)\log(P) - 0.656(\pm0.123), \sigma=0.249 \\
        \nonumber
        M_V=  -3.024(\pm0.097)\log(P) - 0.977(\pm0.111), \sigma=0.225 \\
        \nonumber
        M_I\ =  -3.328(\pm0.098)\log(P) - 1.432(\pm0.108), \sigma=0.217
        \end{eqnarray}

        Therefore, the exclusion or inclusion of \citet{tur02} distances produces almost identical PL relations. These PL relations also agree and are consistent with equations (1)-(3). However, there are some open cluster distances that are not included in \citet{fea99} or T03 (e.g. for Cepheid AQ PUP), and there are certain cases where we have confidence to believe that the distances given in \citet{tur02} and \citet{fea99} are independent (e.g. for Cepheids CEa, CEb and CF CAS). Hence we continue to adopt equations (1)-(3) as the calibrated Galactic PL relations. 

\section{Comparison to published results}


        \begin{table*}
        \centering
        \caption{Comparison of various PL relations$^{\mathrm a}$.}
        \label{tab2}
        \begin{tabular}{lcccccccl} \hline       
        PL relation & $N$       & $a_V$            & $b_V$             & $\sigma_V$ & $a_I$            & $b_I$          & $\sigma_I$ & Ref.$^{\mathrm b}$ \\
        \hline \hline
        GAL-G98     & 28        & $-3.037\pm0.138$ & $-1.021\pm0.040$ & 0.209      & $-3.329\pm0.132$ & $-1.435\pm0.037$ & 0.194 & 1 \\
        GAL-F03     & 32        & $-3.06\pm0.11$   & $-0.989\pm0.034$ & $\cdots$   & $-3.24\pm0.11$   & $-1.550\pm0.034$ & $\cdots$&2\\  
        GAL-T03     & 53        & $-3.141\pm0.100$ & $-0.826\pm0.119$ & 0.24       & $-3.408\pm0.095$ & $-1.325\pm0.114$ & 0.23  & 3\\   
        GAL-Here    & 50        & $-2.999\pm0.097$ & $-0.995\pm0.112$ & 0.226      & $-3.303\pm0.094$ & $-1.450\pm0.108$ & 0.219 & 4\\
        \hline 
        GAL-Theory1 & $\cdots$  & $-2.905$         & $-1.183$         & $\cdots$   & $-3.102$         & $-1.805$         &$\cdots$& 5\\  
        GAL-Theory2 & $\cdots$  & $-2.45\pm0.02$   & $-1.50\pm0.02$   & 0.17       & $-2.78\pm0.01$   & $-2.02\pm0.01$   & 0.13  & 6\\
        GAL-Theory3 & $\cdots$  & $-2.22\pm0.01$   & $-1.62\pm0.01$   & 0.14       & $-2.58\pm0.01$   & $-2.14\pm0.01$   & 0.10  & 7\\
        \hline
        LMC         & $\sim650$ & $-2.760\pm0.030$ & $-1.458\pm0.020$ & 0.160      & $-2.962\pm0.020$ & $-1.942\pm0.010$ & 0.110 & 8 \\ 
        \hline
        \end{tabular}
        \begin{list}{}{}
        \item   $^{\mathrm a}$ $M_{V,I}=a_{V,I}\log(P) + b_{V,I}$, and $\sigma_{V,I}$ is the rms dispersion. For LMC PL relation, assume $\mu_{LMC}=18.50mag.$ 
        \item   $^{\mathrm b}$ Reference: [1] \citet{gie98}; [2] \citet{fou03}; [3] \citet{tam03}; [4] Equations (2) \& (3) from this work; [5] \citet{bar01}, with $Z=0.02$; [6] table 6 from \citet{fio02}, with $Y=0.31$ and $Z=0.02$; [7] table 6 from \citet{fio02}, with $Y=0.28$ and $Z=0.02$; [8] \citet{fre01}. 
        \end{list}
        \end{table*}

\subsection{Comparison to other Galactic PL relations}

        We selected recent empirical PL relations from the literature that give {\it both} V and I band PL relations to be compared with our results, because the PL relations in these two bands are extensively used in extragalactic distance scale studies (e.g., \citealt{fre01}). For example, we exclude the PL relations from \citet{bar03} or \citet{hoy03} because they did not give the I band PL relations in their papers. The selected PL relations, along with our results, are given in Table \ref{tab2}. These include the Galactic PL relations derived by \citet[][GAL-G98]{gie98}, \citet[][GAL-F03]{fou03} and \citet[][GAL-T03]{tam03}. We exclude the {\it Hipparcos}-based Galactic PL relations because these PL relations adopted the slopes from LMC PL relations, and calibrated the zero-points with {\it Hipparcos} data (as those used in, e.g., \citealt{lan99} and \citealt{pat02}). From the table, it can be seen that our results are consistent with the GAL-G98 and GAL-F03 PL relation. However there is some discrepancy between our results and GAL-T03. In this situation, we can use the $t$-statistical test (see, e.g., \citealt{zwi00}) to assess the significance of the difference in slopes under the null hypothesis that the slopes are the same. The results show that the $p$-value for the V and I band slopes are $0.15$ and $0.27$, respectively. Hence the null hypothesis cannot be ruled out at the 95\% confidence level, and our results are also consistent with the GAL-T03 PL relation. 

        The small discrepancy of the PL slopes between our results and T03 is mainly due to the inclusion of additional distance measurements in this work (see Section 2 for details). If we take the arithmetic, unweighted-mean of the 39 Cepheids with $\mu_o(O.C.)$ and $\mu_o(G98)$ in Table \ref{tab1} (as these distance moduli are used in T03), the slopes of the fitted PL relations become steeper: $a_B=-2.750\pm0.123$, $a_V=-3.130\pm0.110$ and $a_I=-3.402\pm0.106$, which agree with the results of T03. However, by taking the weighted-mean distances for these 39 Cepheids, the PL slopes become shallower but still consistent with T03: $a_B=-2.721\pm0.125$, $a_V=-3.102\pm0.112$ and $a_I=-3.373\pm0.109$. Thus the different between our results and T03 is due to the inclusion of additional distance measurements in our study. 
        
        For completeness, we also include the recent theoretical PL relations in Table \ref{tab2} from \citet[][GAL-Theory1]{bar01} and \citet[][GAL-Theory2 \& GAL-Theory3]{fio02} by adopting $Z=0.02$. From the table, our empirical PL relations fairly agree with the theoretical PL relations from \cite{bar01}. However, none of the empirical PL relations given in Table \ref{tab2} agree with the GAL-Theory2 and GAL-Theory3. 

\subsection{Comparison to LMC PL relation}

        Since the PL relation is shown to be different in the LMC and Galaxy by T03 (also in \citealt{fou03} and \citealt{kan03}), we verify this result by comparing our Galactic PL relation to the LMC counterpart, as given in \citet{fre01}. For our Galactic PL relation, the difference in the V and I band slopes is: $\Delta a_V=0.239\pm0.102$ and $\Delta a_I=0.341\pm0.096$, which are $2.3\sigma$ and $3.6\sigma$ results, respectively. We also apply the $t$-statistical test to test for the equality in the slopes of the Galactic and LMC PL relation. The results show that the slopes in the Galactic and LMC PL relation are inconsistent at more than a 95\% confidence level, with $p$-value of $0.017$ and $0.001$ for the V and I band slopes, respectively. Therefore, Cepheids in the Galaxy and the LMC do follow different PL relations, and hence the Cepheid PL relation is not universal.  

\section{The distance to NGC 4258}

        The Galactic PL relation presented in the previous section can be used to find the distance to NGC 4258, because the metallicity in this galaxy is $8.85\pm0.06dex$ \citep[see reference in][]{new01}, which is closer to the Solar value. Furthermore, there is an accurate geometrical distance measurement to NGC 4258 using the water maser in the inner disk of this galaxy \citep{her99}. The measured geometrical distance is $7.2\pm0.5Mpc$, corresponding to $\mu_{geo}=29.28\pm0.15mag.$ 

        There are 15 Cepheids discovered with {\it HST} observations by \citet{new01}, where we adopted the periods and the mean V and I band magnitudes (from ALLFRAME photometry) for these Cepheids. The distance modulus to NGC 4258 can be obtained using the prescription given by \citet{kan03}. We did not include the metallicity correction because it is small \citep{kan03}. The results are given in Table \ref{tab3}, using the different PL relations in Table \ref{tab2}. This shows that the distance to NGC 4258 is consistent with these different empirical Galactic PL relations. However, the distance modulus of $29.49mag.$ (from the PL relations given in equations [2] \& [3]), and the median distance modulus of $29.46mag.$ (from the four empirical Galactic PL relations) is still $\sim1.4\sigma$ and $\sim1.2\sigma$, respectively, away from the water maser distance\footnote{Using the mean magnitudes derived in \citet{kan03} for these 15 Cepheids decreases the $\mu_o$ in Table \ref{tab3} by $\sim0.06mag.$, with a median distance modulus of $29.40mag.$}. Furthermore, all the theoretical PL relations in Table \ref{tab2} give a further and inconsistent distance modulus as compared to the distance moduli obtained from empirical PL relations (see, however, \citealt{cap02} for a way to resolve this discrepancy). This could be due, in part, to the small number of Cepheids discovered in NGC 4258 \citep{new01}. The on-going Cycle 12 {\it HST} observations of NGC 4258 (Program ID: 9810; P.I.: L. Greenhill) that proposed to discover $\sim100$ Cepheids in this galaxy would help to solve the discrepancy between the Cepheid distance and water maser distance.   


        \begin{table}
        \centering 
        \caption{Distances to NGC 4258 with different PL relations$^{\mathrm a}$.}
        \label{tab3}
        \begin{tabular}{lccc} \hline
        PL Relation   &  $\mu_V$ & $\mu_I$ & $\mu_o$ \\
        \hline 
        GAL-G98        & $29.80\pm0.07$ & $29.65\pm0.05$ & $29.43\pm0.06$\\
        GAL-F03        & $29.79\pm0.07$ & $29.66\pm0.05$ & $29.46\pm0.06$\\
        GAL-T03        & $29.73\pm0.07$ & $29.63\pm0.05$ & $29.50\pm0.06$\\
        GAL-Here       & $29.72\pm0.07$ & $29.63\pm0.05$ & $29.49\pm0.06$\\
        \hline
        GAL-Theory1    & $29.80\pm0.07$ & $29.74\pm0.05$ & $29.66\pm0.06$\\
        GAL-Theory2    & $29.57\pm0.06$ & $29.58\pm0.04$ & $29.57\pm0.06$\\
        GAL-Theory3    & $29.41\pm0.06$ & $29.45\pm0.04$ & $29.50\pm0.06$\\
        \hline
        LMC            & $29.90\pm0.07$ & $29.71\pm0.05$ & $29.44\pm0.06^{\mathrm b}$\\
        \hline
        \end{tabular}
        \begin{list}{}{}
        \item   $^{\mathrm a}$ The errors are random error only. The systematic error is about $0.15mag$ \citep{new01}.
        \item   $^{\mathrm b}$ The metallicity correction using the LMC PL relation to this galaxy is $\delta_z=+0.07mag.$ \citep{fre01,new01}. Then the metallicity-corrected distance modulus becomes $29.51\pm0.06mag.$ 
        \end{list}
        \end{table}

        
\section{Conclusion}

        By using the recent independent distance measurements to 50 Galactic Cepheids, we derive a Galactic PL relation in the B, V and I bands. Our analysis differs from that in T03 due to the following aspects: (a) we include other recent independent distance measurements; and (b) we took the weighted-mean to the available distances. The results confirm that the Galactic PL relations are steeper than the LMC counterparts \citep{fou03,tam03}. However, the steepness of the Galactic PL relation is still inconclusive (either as steep as in T03 or as shallow as in this study) because of the small number ($\sim50$) of Cepheids in the sample. Application of the Galactic PL relation from equation (2) \& (3) to determine the distance to NGC 4258 shows that there is still a $\sim1.4\sigma$ discrepancy in distance with the water maser measurements. The Galactic PL relation can be improved if there are more independent distance measurements to Galactic Cepheids in the future, such as the Cycle 12 {\it HST} observations of the nearby Cepheids with astrometric measurements (10 Cepheids from Program 9879, with P.I.: G. F. Benedict, where 3 of them are not included in Table \ref{tab1} [Benedict 2003, private communication]) or the future Space Interferometry Mission (SIM, launch in 2009)\footnote{{\tt http://planetquest.jpl.nasa.gov/SIM/sim\_index.html}}. 


\section*{acknowledgements}
        We would like to thank T. Barnes, F. Benedict, H. Bond, P. Fouqu\'{e}, and G. A. Tammann for useful discussion. Especially thanks to Tammann and anonymous referee for useful suggestions.


\end{document}